\documentclass[a4paper,12pt]{article}
\usepackage{comment}
\usepackage{cite}
\usepackage{amsmath}
\usepackage{amssymb}
\usepackage{amsfonts}
\usepackage[T1]{fontenc}
\usepackage[utf8]{inputenc}
\usepackage{graphicx}
\usepackage{fancyhdr}
\usepackage{float}
\usepackage{xcolor}
\usepackage{authblk}
\usepackage{mathrsfs}
\usepackage{empheq}
\usepackage[hyphens]{url}
\usepackage{hyperref} 
\usepackage[]{breakurl}
\usepackage[]{amsthm}
\usepackage{tikz}
\usetikzlibrary{decorations.markings}

\usepackage{geometry}
\begin{document}

\title{Imaginary part of the conductivity using Kramers-Kronig relations}

\author[$\dagger$]{Jean-Christophe {\sc Pain}\footnote{jean-christophe.pain@cea.fr}\;\; and Mikael {\sc Tacu}\\
\small
$^1$CEA, DAM, DIF, F-91297 Arpajon, France\\
$^2$Universit\'e Paris-Saclay, CEA, Laboratoire Mati\`ere en Conditions Extr\^emes,\\ 
F-91680 Bruy\`eres-le-Ch\^atel, France
}

\date{}

\maketitle

\begin{abstract}
In order to obtain the frequency-dependent photo-absorption in a plasma, both the real and imaginary parts of the AC conductivity are required. The real part can be deduced from the knowledge of the static conductivity (given by the Ziman-Evans formula for instance) and the Drude model. The imaginary part, required for the refraction index, can be obtained using the Kramers-Kronig relations. Usually, it is obtained by complex integration in the complex plane of the usual Kramers-Kronig relations, having $\omega'-\omega$ in the denominator. However, an alternate form of the Kramers-Kronig relation is often used in physics, especially for determining response functions. It has $\omega'^2-\omega^2$ in the denominator. We provide two determinations of the imaginary part of the conductivity for this latter form, one using a decomposition into simple elements, and the other involving a complex integration in a quarter of the complex plane.
\end{abstract}

\section{Introduction}

To determine the frequency-dependent photo-absorption in a plasma, both the real and imaginary components of the AC conductivity are required. The real part can be derived from the static conductivity, obtained, for instance, using the Ziman-Evans formula combined with the Drude model. The imaginary part, essential for calculating the refractive index, can be determined using the Kramers-Kronig relations. Typically, this is done by performing an integration in the complex plane using the standard Kramers-Kronig relation, which has $\omega'-\omega$ in the denominator. However, an alternate form of the Kramers-Kronig relation, commonly used in physics to determine response functions, features $\omega'^2-\omega^2$ in the denominator.

In this work, we present two approaches for computing the imaginary part of the conductivity using the latter formulation. The first method relies on a decomposition into simple elements, while the second involves a complex integration over a quarter of the complex plane.

For the frequency-dependent (or spectral) opacity (photo-absorption cross-section per mass unit) we have (see for instance Ref. \cite{Callaway1974,Johnson2009}):
\begin{equation}
    \kappa(\omega)=\frac{1}{\rho}\frac{4\pi}{n(\omega)c}\Re\left[\sigma(\omega)\right],
\end{equation}
where $\rho$ is the density, $c$ the sppe of light and the refraction index $n(\omega)$ reads
\begin{equation*}
    n(\omega)=\sqrt{\frac{|\epsilon(\omega)|+\Re \left[\epsilon(\omega)\right]}{2}},
\end{equation*}
$\epsilon(\omega)$ being the frequency-dependent dielectric function:
\begin{equation*}
    \epsilon(\omega)=1+4\pi\,i\frac{\sigma(\omega)}{\omega}.
\end{equation*}
Knowing the static (DC) conductivity $\sigma_0$, for instance within the Ziman-Evans approach, the frequency-dependent conductivity can be obtained using the Drude formula:
\begin{equation}
    \sigma(\omega) = \frac{\sigma_0}{(1+\omega^2\tau_p^2)},
\end{equation}
where $\tau_p$ is the relaxation time.

Let $\chi (\omega )=\chi _{1}(\omega )+i\chi _{2}(\omega )$ be a complex function of the complex variable $\omega$, where $\chi _{1}(\omega)$ and $\chi _{2}(\omega)$ are real, and assume that $\chi(\omega)$ is analytic in the closed upper half-plane and vanishes at infinity. The Kramers–Kronig relations read
\begin{equation}
    \chi _{1}(\omega )={\frac{1}{\pi }}{\mathcal {P}}\!\!\int_{-\infty }^{\infty }{\frac {\chi _{2}(\omega ')}{\omega '-\omega }}\,d\omega'
\end{equation}
and
\begin{equation}\label{usualchi2}
    \chi _{2}(\omega )=-{\frac {1}{\pi }}{\mathcal {P}}\!\!\int _{-\infty }^{\infty }{\frac {\chi _{1}(\omega ')}{\omega '-\omega }}\,d\omega ',
\end{equation}
where $\omega$ is real and where $\mathcal {P}$ denotes the Cauchy principal value. 

Such integrals can be calculated by integration in the complex plane using the Cauchy homotopy theorem, stating that 
\begin{equation}
    \oint {\frac {\chi (\omega ')}{\omega '-\omega }}\,d\omega '=0
\end{equation}
for any closed contour within this region. Usually the chosen contour is a closed semi-circle in the upper half-plane, avoiding the pole $\omega '=\omega$.

Thanks to these properties, we can limit the range of integration $[0,+\infty[$. Considering the first relation, giving the real part $\chi_{1}(\omega )$. let us multiply, in the integral, the numerator and denominator of the integrand by $\omega '+\omega$: 
\begin{equation}
    \chi _{1}(\omega )={1 \over \pi }{\mathcal {P}}\!\!\int _{-\infty }^{\infty }{\omega '\chi _{2}(\omega ') \over \omega '^{2}-\omega ^{2}}\,d\omega '+{\omega \over \pi }{\mathcal {P}}\!\!\int _{-\infty }^{\infty }{\chi _{2}(\omega ') \over \omega '^{2}-\omega ^{2}}\,d\omega '.
\end{equation}
Since 
$\chi_{2}(\omega)$ is odd, the second integral vanishes, and we are left with 
\begin{equation}
    \chi_{1}(\omega )={2 \over \pi }{\mathcal {P}}\!\!\int _{0}^{\infty }{\omega '\chi _{2}(\omega ') \over \omega '^{2}-\omega ^{2}}\,d\omega '.
\end{equation}
A similar derivation for the imaginary part gives, {\it mutatis mutandis}:
\begin{equation}
    \chi_{2}(\omega )=-{2 \over \pi }{\mathcal {P}}\!\!\int _{0}^{\infty }{\omega \chi _{1}(\omega ') \over \omega '^{2}-\omega ^{2}}\,d\omega '=-{2\omega \over \pi }{\mathcal {P}}\!\!\int _{0}^{\infty }{\chi _{1}(\omega ') \over \omega '^{2}-\omega ^{2}}\,d\omega '.
\end{equation}
This form of the Kramers–Kronig relations is often used in physics for the derivation of response functions.

Using the Kramers-Kronig formulas, it is possible to get the full complex conductivity. In particular, for the imaginary part of the conductivity, we have, 
\begin{equation*}
\Im\left[\sigma(\omega)\right]=-\frac{2\omega}{\pi}\mathcal {P}\!\!\int _{0}^{\infty}\frac{\Re\left[\sigma(\omega')\right]}{\omega'^{2}-\omega ^{2}}\,d\omega ',
\end{equation*} 
where $\mathcal {P}$ denotes the principal value.

\begin{equation}\label{newchi2}
    \chi _{2}(\omega )=-\frac{2\omega}{\pi}\mathcal{P}\!\!\int_{0}^{\infty}\frac{\chi _{1}(\omega')}{\omega'^{2}-\omega ^{2}}\,d\omega ',
\end{equation} 
where $\chi_1(\omega)=\Re\left[\sigma(\omega)\right]$ and $\chi_2(\omega)=\Im\left[\sigma(\omega)\right]$.
However, such a formula does not lend itself very well to complex integration calculations, on the contrary to Eq. (\ref{usualchi2}). In section \ref{sec2}, we recall the calculation of $\chi_2$ using Eq. (\ref{newchi2}) and the decomposition into simple elements. In section \ref{sec3}, we calculate the integral by complex integration, using a quarter-circle contour, and parts of the $x$ and $y$ axis avoiding a real and an imaginary pole.  

\section{First method: decomposition into simple elements}\label{sec2}

We have to calculate
\begin{equation}
    I=\mathcal {P}\int_0^{\infty}\frac{d\omega'}{(\omega'^2-\omega^2)(1+\tau_p^2\omega'^2)}.
\end{equation}
Using the decomposition into simple elements
\begin{equation}
    \frac{1}{(\omega'^2-\omega^2)(1+\tau_p^2\omega'^2)}=\frac{1}{1+\tau_p^2\omega^2}\left[-\frac{\tau_p^2}{1+\tau_p^2\omega'^2}+\frac{1}{\omega'^2-\omega^2}\right]
\end{equation}
we obtain 
\begin{equation}\label{des}
    I=\frac{1}{1+\tau_p^2\omega^2}\left[-\tau_p^2\int_0^{\infty}\frac{d\omega'}{1+\tau_p^2\omega'^2}+\mathcal{P}\int_0^{\infty}\frac{d\omega'}{\omega'^2-\omega^2}\right].
\end{equation}
We have
\begin{equation}
    \int_0^{\infty}\frac{d\omega'}{1+\tau_p^2\omega'^2}=\left.\frac{1}{\tau_p}\mathrm{arctan}(\tau_px)\right|_0^{\infty}=\frac{\pi}{2\tau_p}.
\end{equation}
The second integral in Eq.(\ref{des}) is
\begin{equation}
    \mathcal{P}\int_0^{\infty}\frac{d\omega'}{\omega'^2-\omega^2}=\frac{1}{\omega}\mathcal{P}\int_0^{\infty}\frac{d\omega'}{\omega'^2-1}
\end{equation}
and
\begin{equation}
    \mathcal{P}\int_0^{\infty}\frac{d\omega'}{\omega'^2-1}=\lim_{\epsilon\rightarrow 0}\left[\int_0^{1-\epsilon}\frac{d\omega'}{\omega'^2-1}+\int_{1+\epsilon}^{\infty}\frac{d\omega'}{\omega'^2-1}\right].
\end{equation}
Making the change of variable $\omega'\rightarrow 1/\omega'$ in the last integral of the right-hand side of the latter equation, we are left with
\begin{equation}
    \mathcal{P}\int_0^{\infty}\frac{d\omega'}{\omega'^2-1}=\lim_{\epsilon\rightarrow 0}\left[\int_0^{1-\epsilon}\frac{d\omega'}{\omega'^2-1}-\int_{0}^{\frac{1}{1+\epsilon}}\frac{d\omega'}{\omega'^2-1}\right],
\end{equation}
or equivalently
\begin{equation}
    \mathcal{P}\int_0^{\infty}\frac{d\omega'}{\omega'^2-1}=\lim_{\epsilon\rightarrow 0}\int_{\frac{1}{1+\epsilon}}^{1-\epsilon}\frac{d\omega'}{\omega'^2-1},
\end{equation}
which tends to 0 as $\epsilon\rightarrow 0$. Finally, we have 
\begin{equation}
    I=-\frac{\tau_p^2}{(1+\tau_p^2\omega^2)}\frac{\pi}{2\tau_p}=-\frac{\pi}{2}\frac{\tau_p}{1+\tau_p^2\omega^2}.
\end{equation}

\section{Second method: complex integration}\label{sec3}

\begin{figure}[!ht]
\begin{center}
\begin{tikzpicture}[scale=1.4,decoration={markings,
      mark=at position 1.5cm with {\arrow[line width=1pt]{>}},
      mark=at position 5cm with {\arrow[line width=1pt]{>}},
      mark=at position 7cm with {\arrow[line width=1pt]{>}},
      mark=at position 15cm with {\arrow[line width=1pt]{>}},
      mark=at position 25cm with {\arrow[line width=1pt]{>}},
      mark=at position 28cm with {\arrow[line width=1pt]{>}},
      mark=at position 30cm with {\arrow[line width=1pt]{>}},
}
]
\draw[help lines,->] (0,0) -- (7,0) coordinate (xaxis);
\draw[help lines,->] (0,0) -- (0,7) coordinate (yaxis);

\node at (3,0) {$\times$};
\node at (3,0.25) {$\omega$};
\node[below] at (0,0) {$O$};

\path[draw,line width=2pt,postaction=decorate] 
(0,0) -- (2.4,0) node[below] {$\omega-\epsilon$} arc (180:0:0.5) node[below] {$\omega+\epsilon$} -- (6,0) node[below] {$R$} arc (0:90:6) -- (0,3) arc (90:-90:0.5) -- (0,0);

\node[below] at (xaxis) {$x$};
\node[left] at (yaxis) {$y$};
\node at (1.5,0.5) [scale=1.5] {$\gamma_1$};
\node at (3,1) [scale=1.5] {$\gamma_2$};
\node at (4.5,0.5) [scale=1.5] {$\gamma_3$};
\node at (4.5,4.75) [scale=1.5] {$\gamma_4$};
\node at (0.5,4) [scale=1.5] {$\gamma_5$};
\node at (0,2.5) {$+i/\tau_p$};
\node at (0.6,1.7) {$i/\tau_p-\epsilon$};
\node at (0.6,3.3) {$i/\tau_p+\epsilon$};
\node at (1,2.5) [scale=1.5] {$\gamma_6$};
\node at (0.5,1) [scale=1.5] {$\gamma_7$};
\end{tikzpicture}
\caption{Contour of integration. 
}\label{fig1}
\end{center}
\end{figure}
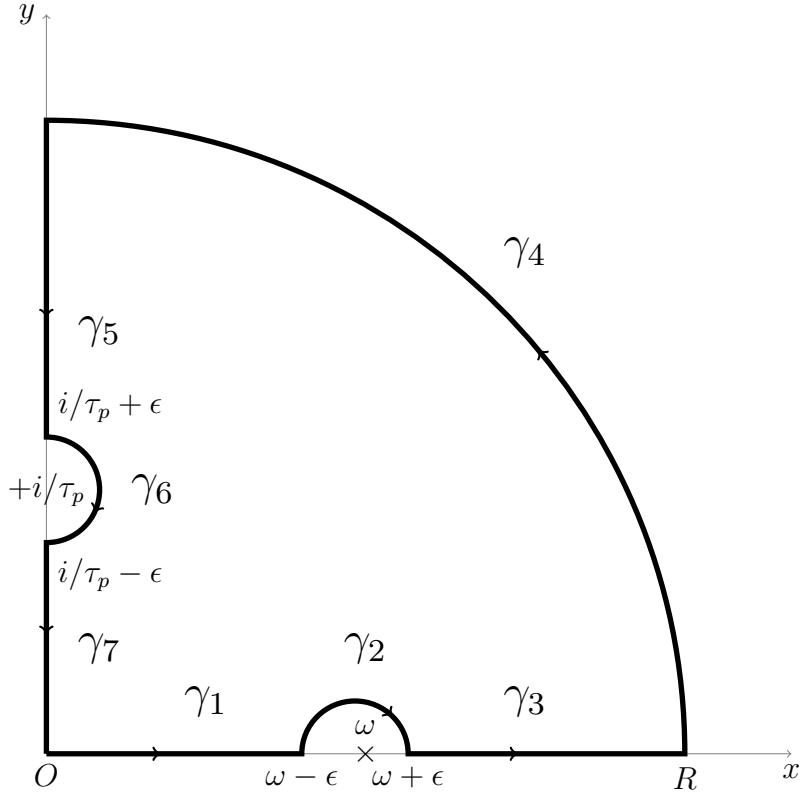

We have
\begin{equation}
    I=\lim_{\epsilon\rightarrow 0}\int_0^{\omega-\epsilon}\frac{d\omega'}{(\omega'^2-\omega^2)(1+\tau_p^2\omega'^2)}+\lim_{\substack{\epsilon\rightarrow 0\\R\rightarrow\infty}}\int_{\omega+\epsilon}^R\frac{d\omega'}{(\omega'^2-\omega^2)(1+\tau_p^2\omega'^2)}.
\end{equation}
Let us set
\begin{equation}
    f(z)=\frac{1}{(z^2-\omega^2)(1+\tau_p^2z^2)}
\end{equation}
and consider the integral in the complex plane, over the path $\gamma=\gamma_1\cup\gamma_2\cup\gamma_3\cup\gamma_4\cup\gamma_5\cup\gamma_6\cup\gamma_7$ represented in Fig. \ref{fig1}. Since the function in the integral is holomorphic over $\mathbb{C}\setminus\left\{\omega, \pm i/\tau_p\right\}$, we have, since $\gamma$ is a closed contour, according to the Cauchy homotopy theorem
\begin{equation}\label{cauchy}
    \oint_{\gamma}f(z)dz=0.
\end{equation}

Equation (\ref{cauchy}) becomes
\begin{align}\label{reli}
    I+&\lim_{\epsilon\rightarrow 0}\int_{\gamma_2}f(z)dz+\lim_{R\rightarrow\infty}\int_{\gamma_4}f(z)dz+\lim_{\substack{\epsilon\rightarrow 0\\R\rightarrow\infty}}\int_{\gamma_5}f(z)dz\lim_{\epsilon\rightarrow 0}\int_{\gamma_6}f(z)dz+\lim_{\epsilon\rightarrow 0}\int_{\gamma_7}f(z)dz=0,
\end{align}
since
\begin{equation}
    I=\lim_{\epsilon\rightarrow 0}\int_{\gamma_1}f(z)dz+\lim_{\substack{\epsilon\rightarrow 0\\R\rightarrow\infty}}\int{\gamma_3}f(z)dz.
\end{equation}
Let us consider the integral over $\gamma_4$ and make the change of variable $z=Re^{i\theta}$:
\begin{equation}
    \int_{\gamma_4}f(z)dz=\int_0^{\pi/2}\frac{iRe^{i\theta}}{(R^2e^{2i\theta}-\omega^2)(1+\tau_p^2R^2e^{2i\theta})}d\theta.
\end{equation}
We have 
\begin{equation}
    \left|\int_{\gamma_4}f(z)dz\right|\leq\int_{0}^{\pi/2}\frac{R}{\left|R^2e^{2i\theta}-\omega^2\right|\left|1+\tau_p^2R^2e^{2i\theta}\right|}d\theta
\end{equation}
and $\left|R^2e^{2i\theta}-\omega^2\right|\geq\left|R^2-\omega^2\right|$ as well as $\left|1+\tau_p^2R^2e^{2i\theta}\right|\geq\left|\tau_p^2R^2-1\right|.$ This leads
\begin{equation}
    \left|\int_{\gamma_4}f(z)dz\right|\leq\int_{0}^{\pi/2}\frac{R}{\left|R^2-\omega^2\right|\left|\tau_p^2R^2-1\right|}d\theta
\end{equation}
i.e.,
\begin{equation}
    \left|\int_{\gamma_4}f(z)dz\right|\leq\frac{\pi}{2}\frac{R}{\left|R^2-\omega^2\right|\left|\tau_p^2R^2-1\right|}
\end{equation}
and the right-hand side tends to 0 when $R\rightarrow\infty$. Thus
\begin{equation}
\int_{\gamma_4}f(z)dz=0.
\end{equation}
As concerns the integral over $\gamma_2$, we make the change of variable $z=\omega+\epsilon e^{i\theta}$ yielding
\begin{equation}
    \int_{\gamma_2}f(z)dz=-\int_0^{\pi}\frac{i\epsilon e^{i\theta}}{((\omega+\epsilon e^{i\theta})^2-\omega^2)(1+\tau_p^2(\omega+\epsilon e^{i\theta})^2)}d\theta,
\end{equation}
which tends to
\begin{equation}
    \int_{\gamma_2}f(z)dz=-i\frac{\pi}{2\omega}\frac{1}{1+\tau_p^2\omega^2}
\end{equation}
when $\epsilon\rightarrow 0$. We have, making the change of variables $z=iy$:
\begin{equation}
    \int_{\gamma_5}f(z)dz=i\int_{1/\tau_p+\epsilon}^{\infty}\frac{dy}{(y^2+\omega^2)(1-\tau_p^2y^2)}.
\end{equation}
and in the same way:
\begin{equation}
    \int_{\gamma_7}f(z)dz=i\int_0^{1/\tau_p-\epsilon}\frac{dy}{(y^2+\omega^2)(1-\tau_p^2y^2)}.
\end{equation}
Using the decomposition in simple elements;
\begin{equation}
    \frac{1}{(y^2+\omega^2)(1-\tau_p^2y^2)}=\frac{1}{1+\tau_p^2\omega^2}\left[\frac{1}{y^2+\omega^2}+\frac{\tau_p^2}{1-\tau_p^2y^2}\right],
\end{equation}
we get
\begin{equation}
    \lim_{\substack{R\rightarrow\infty\\\epsilon\rightarrow 0}}\int_{\gamma_5}f(z)dz+\lim_{\substack{R\rightarrow\infty\\\epsilon\rightarrow 0}}\int_{\gamma_7}f(z)dz=\frac{i\pi}{2\omega}\frac{1}{1+\tau_p^2\omega^2}.
\end{equation}
As concerns the half-circle $\gamma_6$ on the $y$ axis, we make the change of variable $z=\frac{i}{\tau_p}+\epsilon e^{i\theta}$ yielding
\begin{equation}
    \int_{\gamma_6}f(z)dz=-\int_{-\pi/2}^{\pi/2}\frac{i\epsilon e^{i\theta}}{\left[\left(\frac{i}{\tau_p}+\epsilon e^{i\theta}\right)^2-\omega^2\right]\left[1+\tau_p^2\left(\frac{i}{\tau_p}+\epsilon e^{i\theta}\right)^2\right]}d\theta
\end{equation}
which tends to
\begin{equation}
    \int_{\gamma_6}f(z)dz=\frac{\pi}{2}\frac{\tau_p}{\left(1+\tau_p^2\omega^2\right)},
\end{equation}
and thus, according to Eq. (\ref{reli}):
\begin{align}
    I=&-\lim_{\epsilon\rightarrow 0}\int_{\gamma_2}f(z)dz-\lim_{R\rightarrow\infty}\int_{\gamma_4}f(z)dz-\lim_{\substack{\epsilon\rightarrow 0\\R\rightarrow\infty}}\int_{\gamma_5}f(z)dz-\lim_{\epsilon\rightarrow 0}\int_{\gamma_6}f(z)dz-\lim_{\epsilon\rightarrow 0}\int_{\gamma_7}f(z)dz\\
    =&-\frac{\pi}{2}\frac{\tau_p}{1+\tau_p^2\omega^2}+i\frac{\pi}{2\omega}\frac{1}{1+\tau_p^2\omega^2}-i\frac{\pi}{2\omega}\frac{1}{1+\tau_p^2\omega^2},
\end{align}
and finally
\begin{align}
    I=&\mathcal {P}\int_0^{\infty}\frac{d\omega'}{(\omega'^2-\omega^2)(1+\tau_p^2\omega'^2)}=-\frac{\pi}{2}\frac{\tau_p}{1+\tau_p^2\omega^2}
\end{align}
and the imaginary part of the conductivity is thus
\begin{equation*}
\Im\left[\sigma(\omega)\right]=\frac{\omega\tau_p\sigma_0}{1+\tau_p^2\omega^2}.
\end{equation*}   

\section{Conclusion}

As is well known, the calculation of the imaginary part of the AC conductivity can be carried out by a decomposition into simple elements, but the original form of the Kramers-Kronig relations, i.e. with a linear function in the denominator, is more suitable for integration in the complex plane (in the upper half-plane). In this document, we proposed a calculation, by complex integration, of the imaginary part of the AC conductivity based on the second form of the Kramers-Kronig relations, i.e., the one with a quadratic polynomial in the denominator (the second ``physical'' form of the Kramers-Kronig relations). For that purpose, we restricted ourselves to the upper left quadrant with a quarter-circle contour, and found that it is necessary, in this case, to take into account both the real and imaginary poles.

\end{document}